\documentclass[aps,pra]{revtex4}
\usepackage{graphicx}

\begin{document}

\title{Ground state properties of a homogeneous \\
2D system of Bosons with dipolar interactions.}

\author{G.E.~Astrakharchik$^a$, J.~Boronat$^a$, J.~Casulleras$^a$, I.L.~Kurbakov$^b$, Yu.E.~Lozovik$^b$}

\affiliation{$^a$ Departament de F\'{\i}sica i Enginyeria Nuclear, Campus Nord B4-B5, Universitat Polit\`ecnica de Catalunya, E-08034 Barcelona, Spain\\
$^b$ Institute of Spectroscopy, 142190 Troitsk, Moscow region, Russia}

\begin{abstract}
The ground-state phase properties of a two-dimensional Bose system with
dipole-dipole interactions is studied by means of quantum Monte Carlo techniques.
Limitations of mean-field theory in a two-dimensional geometry are discussed. A
quantum phase transition from gas to solid is found. Crystal is tested for existence
of a supersolid in the vicinity of the phase transition. Existence of mesoscopic
analogue of the off-diagonal long-range order is shown in the one-body density
matrix in a finite-size crystal. Non-zero superfluid fraction is found in a
finite-size crystal, the signal being dramatically increased in presence of
vacancies.
\end{abstract}

\maketitle

\section{Model and methods}
We study properties of a two-dimensional (2D) system of bosons with dipolar
interaction. We consider a polarized system and assume that dipolar moments are
oriented perpendicularly to the 2D plane. This assures that the interaction
potential $V_{int}(r)=C_{dd}/|r|^3$ is always repulsive and there are no
instabilities caused by dipolar attraction. The following model Hamiltonian is used
to describe the system:
\begin{eqnarray}
\hat H = -\frac{\hbar^2}{2m}\sum\limits_{i=1}^N\nabla_i+\frac{C_{dd}}{4\pi}\sum\limits_{i<j}^N\frac{1}{|r_i-r_j|^3},
\label{H}
\end{eqnarray}
where $m$ is the dipolar mass and $N$ the number of dipoles.

\begin{figure}
\begin{center}
\includegraphics[width=0.4\columnwidth]{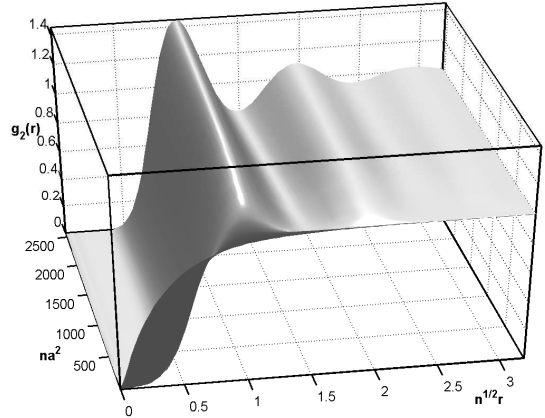}
\caption{Correlation functions in gas phase: Pair distribution function.}
\label{Fig1a}
\end{center}
\end{figure}

Properties of a homogeneous system are governed by one characteristic parameter, the
dimensionless density $nr_0^2$, where the characteristic length $r_0$ is
proportional to the interaction strength: $r_0 = mC_{dd}/4\pi\hbar^2$. Deeply in the
dilute regime one expects that a short-range interaction potential (like the dipolar
one) can be described by only one parameter, namely, the $s$-wave scattering length
$a$. Parameters $r_0$ and $a$ are directly related: $a=e^{2\gamma}r_0=3.17...r_0$.

\begin{figure}
\begin{center}
\includegraphics[width=0.3\columnwidth,angle=-90]{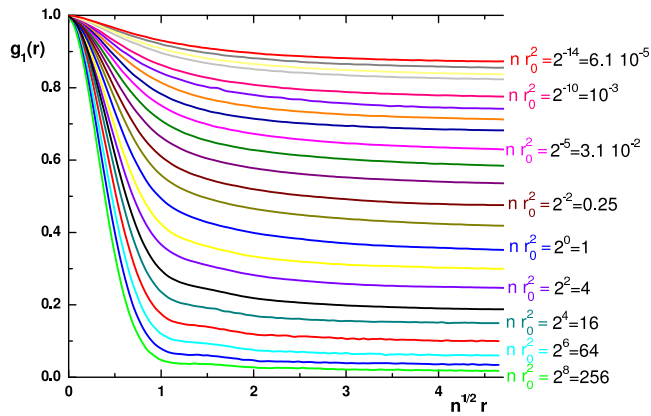}
\caption{Correlation functions in gas phase: One-body density matrix.}
\label{Fig1b}
\end{center}
\end{figure}

We perform a numerical study of the ground-state properties of this system using
quantum Monte Carlo methods. Firstly, using the variational Monte Carlo (VMC) method
it is possible to evaluate multidimensional averages over the trial wavefunction
$\psi_T$. In the calculation of the energy and superfluid fraction $n_S/n$, the
variational parameters in $\psi_T$ are chosen such that they minimize variational
energy. In the calculation of the one-body density matrix we optimize parameters so
that the difference between variational and mixed estimators is minimal. Secondly,
we use the diffusion Monte Carlo (DMC) method based on solving Schr\"odinger
equation in imaginary time at $T=0$. The DMC method permits to find the ground
state energy $E$ of a bosonic system exactly (in statistical sense). Also superfluid
density $n_S/n$ and local quantities (e.g. $g_2(z)$, $S_k$, {\it etc.}) can be found
in a ``pure'' (non-depending on the choice of trial w.f.) way. An extrapolation
procedure can be used for predictions of non-local quantities (e.g. $g_1(z)$, $n_k$,
{\it etc.}).

\section{Trial wave function}
We construct the trial wave function (w.f.) in the following form:
\begin{eqnarray}
\psi_T(r_1,...,r_N) = \prod_{i<j}^Nf_2(|r_i-r_j|)\times
\prod_{k=1}^M\left(\sum_{l=1}^N \exp\{-\alpha(r_l-r_k^{latt.})^2\}\right)
\label{psi}
\end{eqnarray}
where $r_i, i=\overline{1,N}$ are particle coordinates and $r_k^{latt.},
k=\overline{1,M}$ are coordinates of triangular lattice sites. Trial w.f.
(\ref{psi}) is symmetric under exchange of any two particles. The two-body Jastrow
term $f_2(r)$ is chosen\cite{Astrakharchik07a,Astrakharchik07} at short distances as
a solution of the 2-body scattering problem at zero energy. At large distances, the
functional form of the hydrodynamic solution is used \cite{Reatto67}. Thus, $f_2(r)$
accounts for pair-collisions relevant for short distances and collective behavior
(sound) at large distances. One should note that $\psi_T$ is not of a
Nosanow-Jastrow type, as by moving one particle all $M$ terms in the product in
(\ref{psi}) are changed, thus introducing a global change ({\it i.e.} depending as
well on coordinates of other particles), so that this term is not a one-body term,
but rather a many-body term. Another feature of this w.f. is that number of the
particles $N$ can be different from number of lattice sites $M$ making it suitable
for studying of a crystal with vacancies.

The parameter $\alpha$ describes particle localization close to lattice sites. Typical
dependence (at sufficiently large density, $nr_0^2\gtrsim 10$) of variational energy
on parameter $\alpha$ shows two minima. Position of the first minimum is $\alpha=0$.
In this case, the translational invariance is preserved and density profile is flat.
This minimum corresponds to a gas/liquid state. In the second minimum $\alpha$ is
finite and translational invariance is broken. Density profile has crystal symmetry.
This minimum corresponds to a solid state. Thus, with the single trial w.f.
(\ref{psi}) and different variational parameters we are able to describe distinct
phases.

\section{Results}
\subsection{Quantum phase transition}
The ground state phase at small densities corresponds to a gas, solid being
metastable. On the contrary, at large densities solid state is energetically
preferable. The critical density of the quantum phase transition $n_cr_0^2=290(30)$
was obtained by constructing fits to the energy of gas and solid phases
\cite{Astrakharchik07}. This estimation of the critical density is in agreement with
a Path Integral Monte Carlo calculation\cite{buchler07} done at low finite
temperature $n_cr_0^2=320(140)$. Green's Function Monte Carlo
calculation\cite{Mora07} provided a slightly lower critical density
$n_cr_0^2=230(20)$ but in this case a discrete model was used, and by decreasing the
filling factor, a small increase in the critical value was found.

Fig.~\ref{Fig1a} shows the pair distribution function
$g_2(r)=\langle\Psi^\dagger(0)\Psi^\dagger(r)\Psi(r)\Psi(0)\rangle/n^2$ in the gas
phase in a wide range of densities. At the largest density (close to the phase
transition) there is a well pronounced first peak followed by a number of well
visible oscillations. This is a manifestation of strong correlations present in the
system close to the point of a quantum phase transition. As the density is lowered
the height of the peak is decreased and eventually it disappears, leading to a
smooth behavior without any visible oscillations. This smooth behavior is
characteristic for weakly-interacting Bose systems.

\subsection{Dilute regime}
In the dilute regime one expects the mean-field theory to be applicable. As was
rigorously derived in Ref.~[\cite{Lieb01}], the 2D Gross-Pitaevskii equation (GPE) has
a coupling constant $g_{2D}=4\pi\hbar^2/m|\ln(na^2)|$ dependent on the density. This
leads to a logarithmic dependence of the mean-field energy on the
density\cite{Schick71}
\begin{eqnarray}
\frac{E_{MF}}{N}=\frac{1}{2}g_{2D}n=\frac{2\pi\hbar^2}{m|\ln(na^2)|}
\label{E}
\end{eqnarray}
It turns out that the mean-field contribution to the energy (\ref{E}) is the only
well established term. We perform a study of beyond mean-field terms. Comparison to
numerical results for hard-disks\cite{Pilati05} show that for densities
$na^2\lesssim 10^{-6}$ the exact shape of the interaction potential is no longer
important, and that the only relevant parameter is $s$-wave scattering length. The
peculiarity of a two-dimensional system is that for such a small densities,
$10^{-50}<na^2< 10^{-9}$, there is a notable difference (of several percent) between
MF-GPE and exact result. Moreover, to our knowledge there is no analytical theory,
able to reproduce correctly the energy in the whole region of the universal regime.
Summarizing, the mean-field description has limitations (failure) in a
two-dimensional system in the universal regime.

The one-body density matrix $g_1(r)=\langle\Psi^\dagger(r)\Psi(0)\rangle/n$ in the
gas phase has a finite asymptotic value, as reported in Fig.~\ref{Fig1b}. In the
thermodynamic limit ($N\to\infty$) finite asymptotic values are manifestations of
off-diagonal long-range order (ODLRO). The asymptotic value gives the condensate
fraction. In the dilute regime almost all of the particles are condensed, but
increasing the density the stronger interactions deplete the condensate and
condensation fraction drops down to $~1\%$ close to the phase transition point. An
important question is what happens to the condensate as system crystallizes.

\begin{figure}
\begin{center}
\includegraphics[width=0.3\columnwidth,angle=-90]{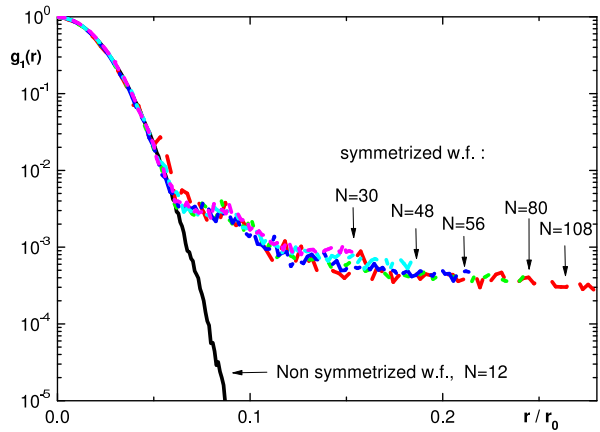}
\caption{Solid phase, $nr_0^2=290$: one-body density matrix for different system
sizes. Solid line --- non symmetrized w.f.}
\label{Fig2a}
\end{center}
\end{figure}

\subsection{Study of a supersolid}
There are several definitions of a {\em supersolid}:
\begin{enumerate}
\item
{\em Spatial order of a solid + finite superfluid density}\\
\item
{\em Spatial order of a solid (broken-symmetry oscillations in diagonal element of OBDM)
+ off-diagonal long-range order in OBDM}
\end{enumerate}
Generally, it is believed that both definitions are equivalent.

Reduced dimensionality increases the role of quantum fluctuations. This makes a
two-dimensional crystal a good candidate for having a supersolid. Notice that in a
one-dimensional, system quantum fluctuations destroy crystalline long-range diagonal
order. A previous study for the presence a supersolid in a two-dimensional dipolar
system is not conclusive. Low-temperature (PIMC) simulation\cite{buchler07} shows
that gas phase is completely superfluid, while no superfluid fraction is found in
crystal phase. Still, the presence of (a possible) supersolid can be masked by much
smaller critical temperature in a crystal. A zero-temperature method was used with a
symmetrized trial w.f. in [\cite{Mora07}]. No conclusions were drawn for
presence/absence of a supersolid due to an unsufficient overlap of the trial w.f. with
the actual ground state.

Fig.~\ref{Fig2a} shows the one-body density matrix $g_1(r)$ in the crystal phase close to
the phase transition. While energy in a crystal is not sensitive to symmetrization
of the wavefunction, it is crucial to symmetrize w.f. in the calculation of $g_1(r)$.
Indeed, without symmetrization off-diagonal element $g_1(r)$ decays exponentially
fast to zero, see thick line in Fig.~\ref{Fig2a}. Using symmetrized w.f. we find
instead a finite asymptotic value (of the order of $3\times 10^{-4}$ for $N=108$
particles). There is a certain decay of the condensate fraction as system size
increases.

Finite-size effects are very important in a $2D$ dipolar system. Indeed, the
characteristic dependence of the energy, OBDM limiting value, {\it etc.} is
$1/\sqrt{N}$ instead of $1/N$ in short-range potentials as can be seen from the tail
correction of the potential energy. For this reason a proper study of the supersolid
in the thermodynamic limit should be done. Here, we limit ourselves to some
preliminary results for a finite-size system.

\begin{figure}
\begin{center}
\includegraphics[width=0.3\columnwidth,angle=-90]{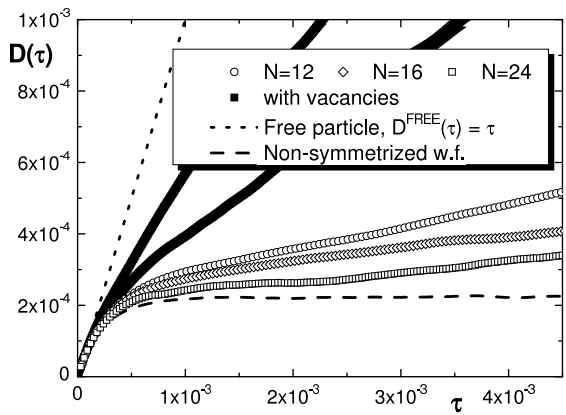}
\caption{Solid phase, $nr_0^2=290$: winding number, open symbols: $N=M=12$
(circles), $16$ (diamonds), $24$ (squares); solid symbols $M=24,N=23$ (lower curve),
$M=24;N=22$ (upper curve); dashed line --- non symmetrized w.f; dotted line ---
diffusion constant of a free particle.}
\label{Fig2b}
\end{center}
\end{figure}

The superfluid fraction corresponds to the slope of the winding number (diffusion
coefficient $D$ of the center of masses in imaginary time $\tau$). Once again,
symmetrization is crucial, otherwise artificial zero slope is obtained (dashed line
in Fig.~\ref{Fig2b}). Using properly symmetrized w.f. we find superfluid signal. The
signal gets weaker as we increase number of particles $N$ (see open symbols in
Fig.~\ref{Fig2b}).

The trial w.f. (\ref{psi}) permits us to investigate role of vacancies. A system
with $M=24$ lattice sites and $0; 1; 2$ vacancies is studied. The superfluid density
experiences dramatic effects in the presence of vacancies. The signal increases from
$\approx 2\%$ for 0 vacancies to $\approx 40\%$ for 2 vacancies.

\section{Conclusions}
To summarize, the Diffusion Monte Carlo method was used to study the properties of a
dipolar 2D Bose system at $T=0$. The ground state energy, pair distribution
function, one-body density matrix were calculated in a wide range of densities. The
gas-solid quantum phase transition is found at density $nr_0^2 = 290(30)$.
Limitations (failure) of mean-field description were pointed out in the universal
low-density regime. Existence of mesoscopic analogue of the off-diagonal long-range
order was shown in one-body density matrix in a finite-size crystal close to phase
transition. Non-zero superfluid fraction was found in a finite-size crystal. The
superfluid signal is dramatically increased in presence of vacancies.


\end{document}